\documentclass[a4paper,12pt,twoside,final]{article}

\usepackage[in,plain]{fullpage}

\usepackage[latin1]{inputenc}

\usepackage{amsmath,amssymb,bm,graphicx}

\usepackage{mathptmx}

\usepackage[mathscr]{euscript}

\usepackage[small,bf]{caption}

\usepackage[colorlinks,linkcolor=blue,citecolor=blue,urlcolor=blue]{hyperref}
\makeatletter
\renewcommand\@makefntext[1]{\leftskip=0.0em\hskip-0.5em\@makefnmark#1}
\makeatother
\renewcommand\dots{\relax\ifmmode\ldots\else$\,\ldots\,$\fi}

\def\rv{\textrm{r.\,v.}}

\def\pdf{\textrm{p.\,d.\,f.}}

\def\cauchy#1#2{\ensuremath{\mathrm{Cauchy}\mkern1mu(#1,#2)}}

\def\leq{\leqslant}
\def\geq{\geqslant}
\def\abs#1{\ensuremath{\lvert\,#1\,\rvert}}

\begin{document}

\renewcommand{\thefootnote}{\fnsymbol{footnote}}

\begin{titlepage}

\thispagestyle{empty}

\begin{center}

{\Large\textbf{Empirical scaling of the length of the longest \\ increasing subsequences of random walks}}

\vspace{6ex}

{\large\textbf{J. Ricardo G. Mendon\c{c}a}\footnote{Email: {\tt\href{mailto:jricardo@usp.br}{\nolinkurl{jricardo@usp.br}}}.}}

\vspace{2ex}

\textit{\mbox{Escola de Artes, Ci\^{e}ncias e Humanidades, Universidade de S\~{a}o Paulo} \\ \mbox{Rua Arlindo Bettio 1000, Ermelino Matarazzo, 03828-000 S\~{a}o Paulo, SP, Brazil}}

\vspace{4ex}

{\large\textbf{Abstract}}

\vspace{2ex}

\parbox{385pt}
{We provide Monte Carlo estimates of the scaling of the length $L_{n}$ of the longest increasing subsequences of $n$-steps random walks for several different distributions of step lengths, short and heavy-tailed. Our simulations indicate that, barring possible logarithmic corrections, $L_{n} \sim n^{\theta}$ with the leading scaling exponent $0.60 \lesssim \theta \lesssim 0.69$ for the heavy-tailed distributions of step lengths examined, with values increasing as the distribution becomes more heavy-tailed, and $\theta \simeq 0.57$ for distributions of finite variance, irrespective of the particular distribution. The results are con\-sis\-tent with existing rigorous bounds for $\theta$, although in a somewhat surprising manner. For random walks with step lengths of finite variance, we conjecture that the correct asymptotic behavior of $L_{n}$ is given by $\sqrt{n}\ln n$, and also propose the form of the subleading asymptotics. The distribution of $L_{n}$ was found to follow a simple scaling form with scaling functions that vary with $\theta$. Accordingly, when the step lengths are of finite variance they seem to be universal. The nature of this scaling remains unclear, since we lack a working model, microscopic or hydrodynamic, for the behavior of the length of the longest increasing subsequences of random walks.

\vspace{4ex}

{\noindent}\textbf{Keywords}: \mbox{LIS} $\cdot$ \mbox{correlated random variables}  $\cdot$ \mbox{heavy tail} $\cdot$ \mbox{random walk} $\cdot$ \mbox{time series} $\cdot$ \mbox{universality}

\vspace{2ex}

{\noindent}\textbf{PACS}: 02.50.-r $\cdot$ 05.40.Fb $\cdot$ 05.45.Tp

\vspace{2ex}

{\noindent}\textbf{MSC 2010}: 60G50 $\cdot$ 60G51 $\cdot$ 82C41

\vspace{2ex}

{\noindent}\textbf{Journal ref.}: \href{http://dx.doi.org/10.1088/1751-8121/aa56a3}{\textit{J.\ Phys.\ A: Math.\ Theor.} \textbf{50}\,(8), 08LT02 (2017)}}

\end{center}

\end{titlepage}


\section{\label{intro}Introduction}

The longest increasing subsequence (LIS) problem is to find an (weakly or strictly) increasing subsequence of maximum length of a given finite sequence of $n$ elements taken from a partially ordered set. The most venerable problem of this kind is that of determining the LIS of a random permutation. The problem seems to have been first posed by S.~Ulam in the early 1960s (apparently motivated by the sorting of bridge hands), who also predicted, based on Monte Carlo simulations for $4 \leq n \leq 10$, that the expected length $L_{n}$ of the LIS of random permutations converges like $\mathbb{E}(L_{n})/\sqrt{n} \to c \simeq 1.7$, and that the distribution of $L_{n}$ should be normal \cite{ulam}. Subsequent numerical and analytical work showed that $\lim_{n \to \infty}\mathbb{E}(L_{n})/\sqrt{n}$ indeed exists and that $c=2$ exactly \cite{baer,hammersley,vershik,logshepp}, but larger Monte Carlo simulations and asymptotics indicated significant deviations from normality \cite{odlyzko}. The complete resolution of the LIS problem for random permutations conflated approaches from diverse and seemingly unrelated fields of mathematics and physics, culminating with the exact determination of the full distribution of the (properly scaled) fluctuations of $L_{n}$ about the $2\sqrt{n}$ limit as the Tracy-Widom distribution for the fluctuations of the largest eigenvalue of a random GUE matrix about the soft edge of the spectrum \cite{t-widom,bdj99}. For comprehensive expositions and further references on the LIS problem for random permutations we refer the reader to \cite{patience,deift,moerbeke,romik}.

Recently, another incarnation of the LIS problem has been posed: what is the behavior of the LIS of a random walk? In \cite{angel,pemantle}, the authors showed that, barring logarithmic corrections, the expected length of the LIS of a random walk on the real line when the step lengths have zero mean and finite positive variance scales with the length $n$ of the walk roughly like $\mathbb{E}(L_{n}) \sim \sqrt{n}$, while the length of the LIS of heavy-tailed random walks with step lengths of infinite variance scales like $\mathbb{E}(L_{n}) \sim n^{\theta}$ with an exponent between $0.690$ and $0.815$. Besides these bounds on $\theta$, not much is known about the LIS of random walks.

In this paper we investigate the scaling behavior of the LIS of random walks by Monte Carlo simulations to provide estimates for the exponent $\theta$ together with the empirical probability distribution of $L_{n}$ for a couple of different distributions of step lengths, namely, the uniform, Laplace (double exponential) and Gaussian distributions, that have finite positive variance, and the symmetric $\alpha$-stable distributions with characteristic exponents $\alpha=\frac{1}{2}$, $1$, $\frac{3}{2}$ and $\frac{7}{4}$, that have infinite variance (and for $\alpha \leq 1$ also have infinite mean) \cite{stable}.


\section{\label{lisrw}Longest increasing subsequences of random walks}

\begin{figure}
\centering
\includegraphics[viewport=30 65 550 245, scale=0.60, clip]{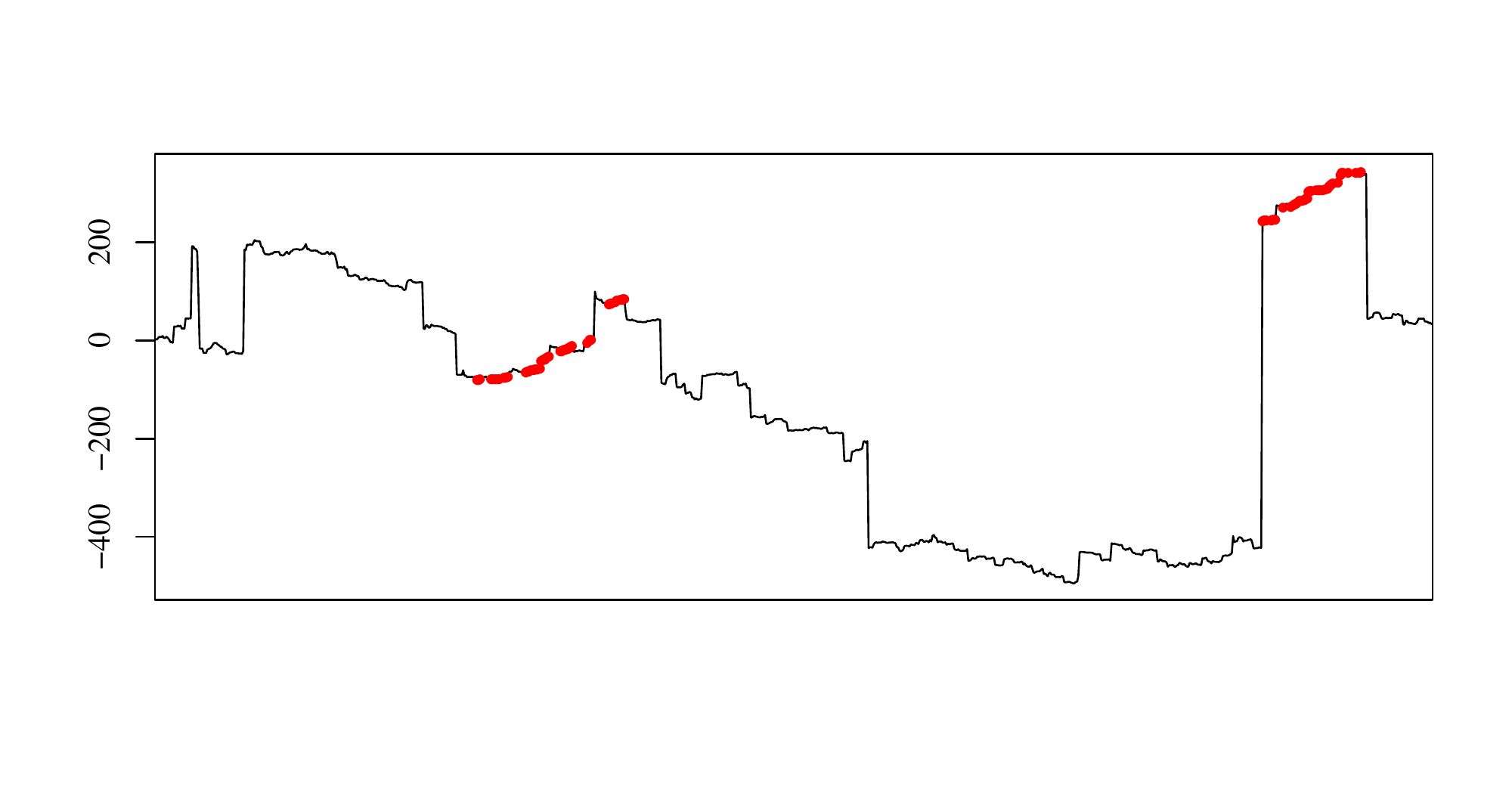} \\[1ex]
\includegraphics[viewport=30 65 550 245, scale=0.60, clip]{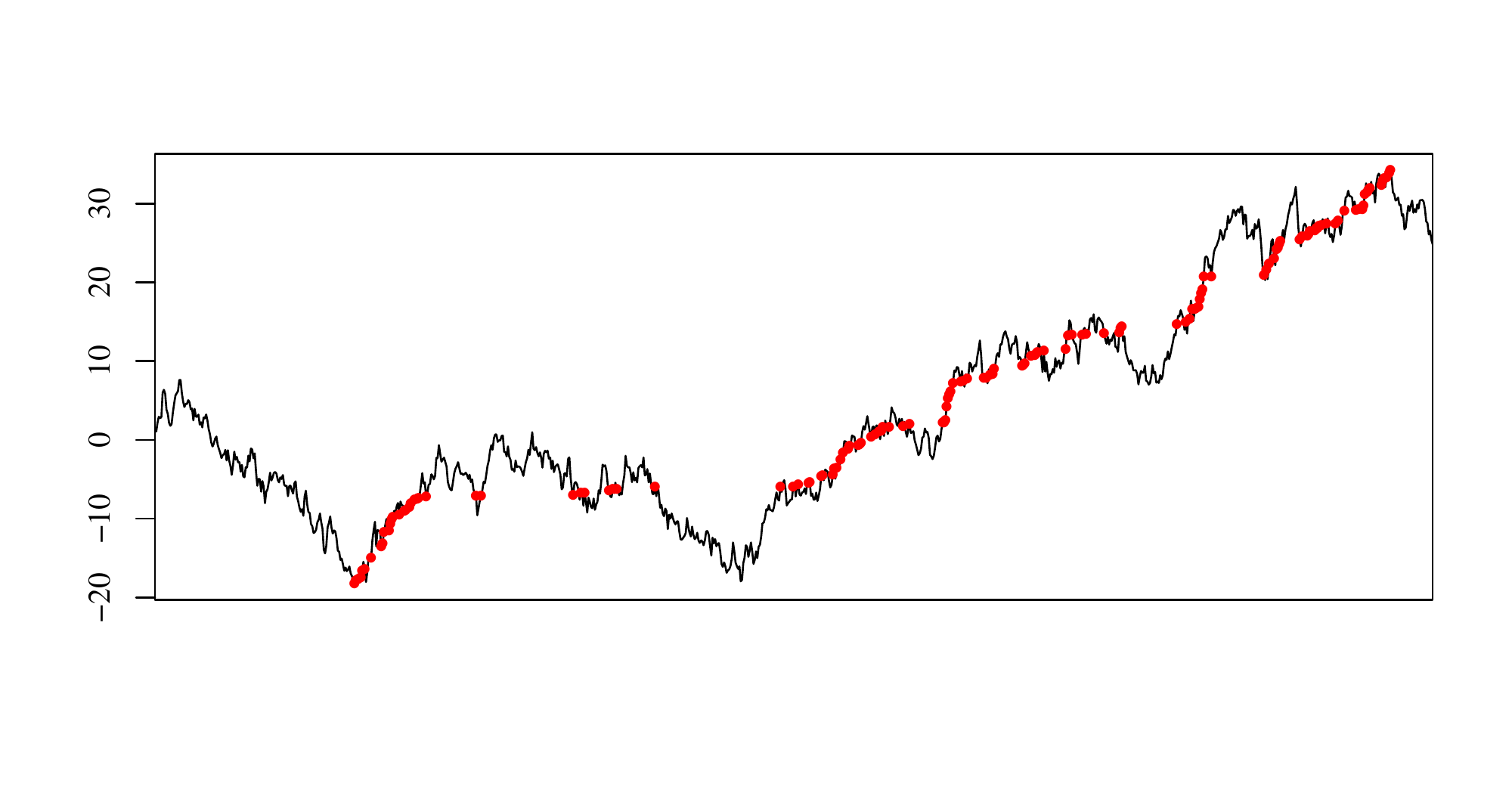}
\caption{\label{fig:rwrw}A \cauchy{0}{1} (top) and a Gaussian $N(0,1)$ (botton) random walk of $1000$ steps each together with one of their longest increasing subsequences (red dots). In these examples, $L_{n}=89$ for the Cauchy random walk and $L_{n}=138$ for the Gaussian random walk. These values for $L_{n}$ are atypical, since random walks with increments of infinite variance have in general longer LIS than those with increments of finite variance.}
\end{figure}

Let $\mathscr{S}_{n}=(S_{1}, \dots, S_{n})$ be the sequence of subsequences of a random walk
\begin{equation}
\label{eq:rws}
S_{n}=X_{1}+\cdots+X_{n}
\end{equation}
of length $n$ with the $X_{i}$, $i=1, \dots, n$, independent random variables (\rv's) identically dis\-trib\-uted according to some continuous probability distribution function (\pdf). The sequence $\mathscr{S}_{n}$ can be though of as a time-series of correlated \rv's, since each term $S_{i}=S_{i-1}+X_{i}$, $i=1, \dots, n$ ($S_{0}=0$). The longest increasing subsequence of $\mathscr{S}_{n}$ is the longest subsequence $S_{i_{1}} \leq S_{i_{2}} \leq \cdots \leq S_{i_{L}}$ of $\mathscr{S}_{n}$ such that $1 \leq i_{1} < i_{2} < \cdots < i_{L} \leq n$, with $L$ the length of the LIS. Note that there may be more than one `longest' increasing subsequence for a given $\mathscr{S}_{n}$; in fact, there may be as many as $O(2^{\lfloor n/2 \rfloor})$ increasing subsequences of length $\lceil n/2 \rceil$ each---think, for example, of the sequence $(2,1,4,3,6,5,\dots)$, ending with $(\cdots, n, n-1)$ if $n$ is even or $(\cdots, n-2, n)$ if it is odd--- but we will be concerned only with their length. Note also that since the $S_{i}$ are continuous \rv's, the distinction between weakly and strictly increasing subsequences is immaterial. Figure~\ref{fig:rwrw} displays a Cauchy and a Gaussian random walk, for which the step lengths follow, respectively, a \cauchy{0}{1} and a normal $N(0,1)$ distribution, together with one LIS each. The \cauchy{0}{1} distribution is but the symmetric $\alpha$-stable distribution with characteristic exponent $\alpha=1$ \cite{stable}, and henceforth we refer to the $\alpha=1$ stable distribution and to the Cauchy distribution interchangeably.

One can reasonably expect that the leading asymptotic behavior of the length of the LIS of a random walk scales with the length of the walk as
\begin{equation}
\label{eq:asymp}
L_{n} \sim cn^{\theta}
\end{equation}
for some positive constant $c$ and $\frac{1}{2} \leq \theta \leq 1$, where the lower bound stems from the Erd\H{o}s-Szekeres theorem \cite{erdosze} and the upper bound is obvious. More refined information on the scaling of $L_{n}$, however, was obtained only recently. In \cite{angel}, the authors showed that when the distribution of the step lengths of the random walk has zero mean and finite positive variance, then for all $\epsilon > 0$ and large enough $n$ the length $L_{n}$ of the LIS of $\mathscr{S}_{n}$ observes
\begin{equation}
\label{eq:finite}
c\sqrt{n} \leq \mathbb{E}(L_{n}) \leq n^{\frac{1}{2}+\epsilon}
\end{equation}
for some positive constant $c$. The upper bound in (\ref{eq:finite}) does not preclude a logarithmic correction, to the effect that it could actually be read like $\mathbb{E}(L_{n}) \leq \sqrt{n}(\ln n)^{a}$ for some $a \geq 0$ and, in fact, whether there is such a logarithmic correction is an open question. It should be remarked that for a simple random walk (steps $\pm 1$) on $\mathbb{Z}$, $\mathbb{E}(L_{n}') \geq c\sqrt{n}\ln n$, where $L_{n}'$ is the length of the \textit{weakly} increasing subsequence, and that the arguments leading to this bound seem to be valid also in the more general case of integer-valued (zero mean, finite variance) random walks on $\mathbb{Z}$ (for example, steps $\pm 1$, \dots, $\pm k$, $k$ finite) \cite{angel}. We are, however, interested in random walks on $\mathbb{R}$, for which the current rigorous bounds read (when the step lengths have zero mean and finite variance) like (\ref{eq:finite}). When the distribution of the increments of the random walk has infinite variance, otherwise, it has been shown that $L_{n}$ behaves like~\cite{pemantle}
\begin{equation}
\label{eq:infinite}
n^{\beta_{0}-o(1)} \leq \mathbb{E}(L_{n}) \leq n^{\beta_{1}+o(1)},
\end{equation}
with $\beta_{0} = 1+W_{0}(-\frac{1}{4}\ln 2)/\ln 2 = 0.690\,093\cdots$ the positive solution of $x+2^{-1-x}=1$, where $W_{0}(z)$ is the principal, upper branch of the Lambert $W$ function \cite{lambert}, while $\beta_{1} = 0.814\,834\cdots$ is obtained from the numerical solution of an implicit expression involving a non-elementary integral. Neither $\beta_{0}$ nor $\beta_{1}$ are sharp. These bounds were obtained for a somewhat contrived `fat-tailed' random walk on a non-Archimedean totally ordered space that, however, behaves like an ultra-heavy tailed $\alpha$-stable random walk with $\alpha=0$. It is also known that the exponent $\theta$ must be strictly greater than $\frac{1}{2}$ for symmetric $\alpha$-stable distributions of step lengths with small enough $\alpha$ \cite{pemantle}.


\section{\label{mcarlo}Scaling behavior and empirical distribution}

For each given distribution of step lengths and length $n$ of the random walk, we generate $10^{4}$ realizations of $\mathscr{S}_{n}$, compute the sample mean and variance of the length $L_{n}$ of the LIS of the $\mathscr{S}_{n}$, and analyze these quantities as a function of $n$ and the underlying distribution of step lengths. In our simulations $n$ range from $10^{4}$ to $10^{8}$ and the distributions investigated are the uniform $H(x+\frac{1}{2})-H(x-\frac{1}{2})$, where $H(x)$ is the Heaviside step function, Laplace (double exponential) $\frac{1}{2}\exp(-\abs{x})$, and Gaussian $\exp(-\frac{1}{2}x^{2})/\sqrt{2\pi}$ distributions, all with zero mean and finite positive variance, and the standard symmetric $\alpha$-stable distributions with characteristic exponents $\alpha=\frac{1}{2}$, $1$, $\frac{3}{2}$ and $\frac{7}{4}$, that are heavy-tailed with tails proportional to $\abs{x}^{-1-\alpha}$ and possess infinite variance \cite{stable}.


\subsection{Leading scaling exponent}

\begin{figure}[t]
\centering
\includegraphics[viewport=0 10 480 460, scale=0.38, clip]{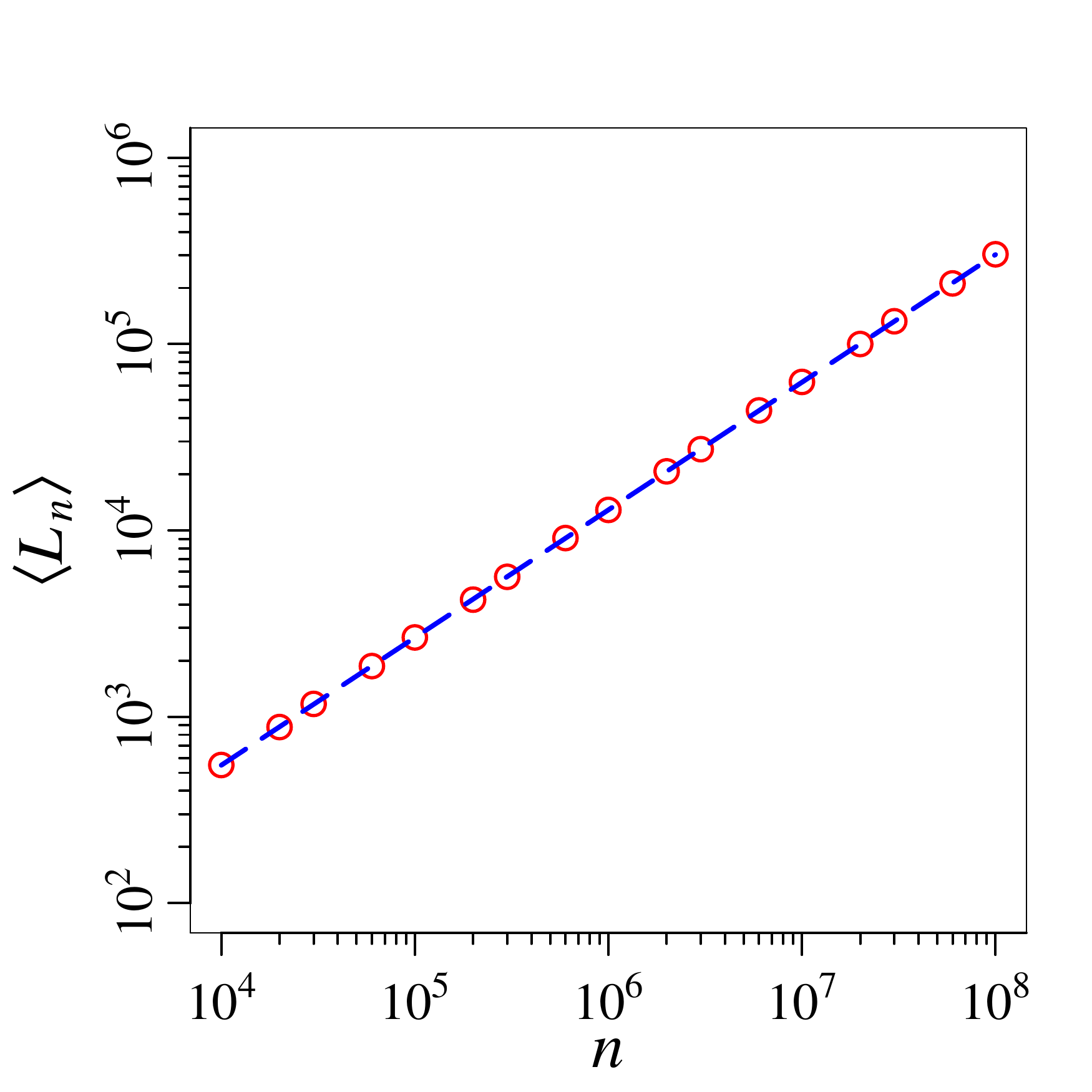} \hspace{1em}
\includegraphics[viewport=0 10 480 460, scale=0.38, clip]{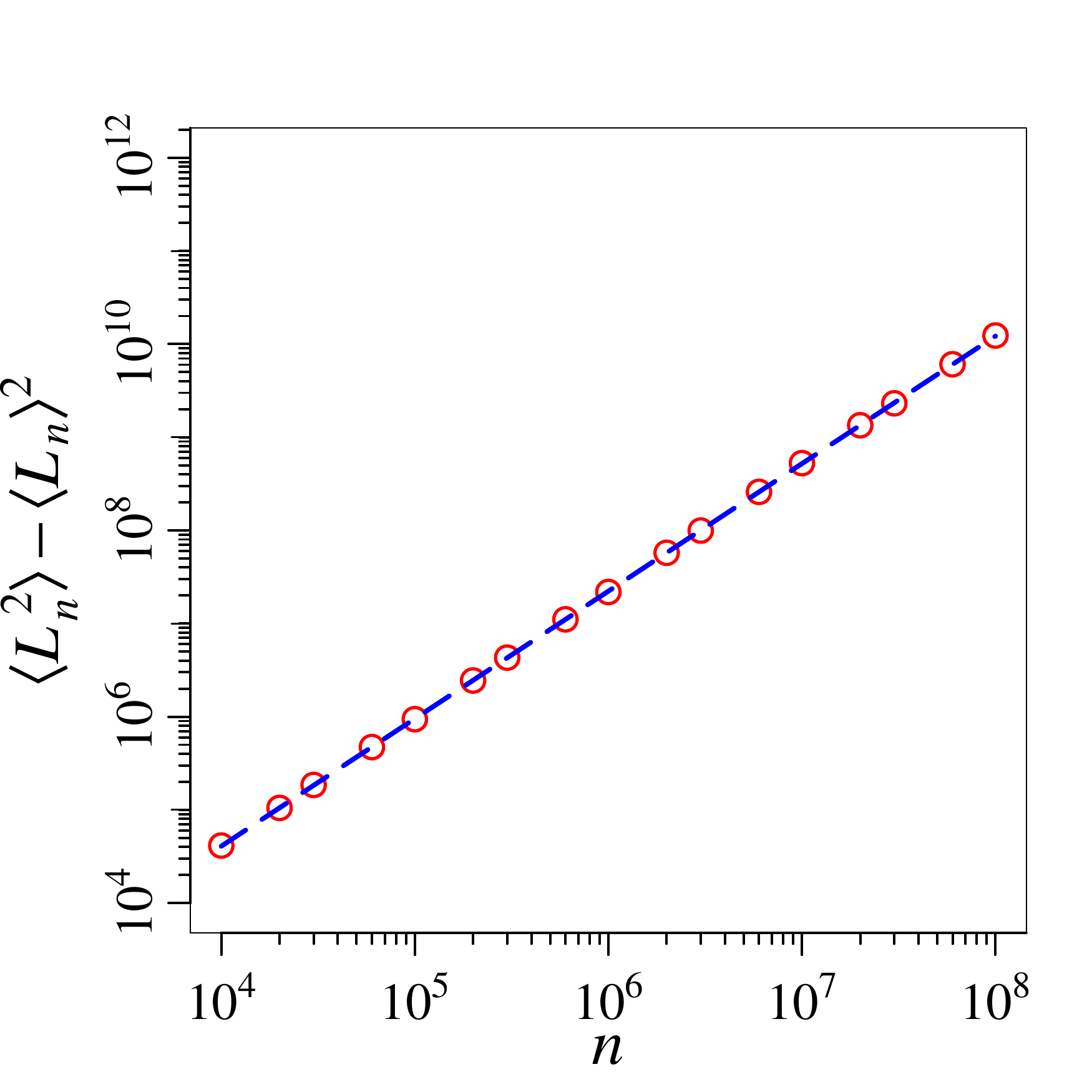}
\caption{\label{fig:theta}Log-log plots of the empirical mean (left panel) and variance (right panel) of $L_{n}$ for Cauchy random walks together with least-squares fits (dashed lines). Each point corresponds to an average over $10^{4}$ sample random walks.}
\end{figure}

We briefly detail the analysis for Cauchy random walks, that is then repeated for the other distributions and summarized in table~\ref{tab:summary}. Figure~\ref{fig:theta} displays log-log plots of $\langle L_{n} \rangle$ and $\langle L_{n}^{2}\rangle-\langle L_{n} \rangle^{2}$ versus $n$ for Cauchy random walks. The plots depict impressive straight lines over four decades of data. Least-squares fits provide the estimates
\begin{subequations}
\begin{equation}
\label{eq:theta}
\langle L_{n} \rangle = 0.999(4)\,n^{\theta}, \quad \theta = 0.6851(3),
\end{equation}
\begin{equation}
\label{eq:gamma}
\langle L_{n}^{2}\rangle-\langle L_{n} \rangle^{2} = 0.1365(17)\,n^{2\gamma}, \quad 2\gamma = 1.3689(9),
\end{equation}
\end{subequations}
where the numbers between parentheses indicate the uncertainty in the last digit(s) of the data. At first it may come as a surprise that the above estimate for $\theta$ approaches the rigorous lower bound in (\ref{eq:infinite}) from below. Note, however, that the lower bound $\beta_{0}$ was obtained for a symmetric `fat-tailed' random walk, which can be thought of as equivalent to an $\alpha \to 0$ stable random walk, and it is not clear whether the same bounds should hold in both cases. Moreover, since $\beta_{0}$ is not sharp for `fat tails', the $o(1)$ term in (\ref{eq:infinite}) may just be showing up in the data. The exponent $\theta$ is, nevertheless, expected to approach the fat-tail exponent as $\alpha \to 0$, and, indeed, as we see from table~\ref{tab:summary}, our data indicate that $\theta$ approaches $\beta_{0}$ as $\alpha \to 0$, settling most likely above it and within the bounds provided by (\ref{eq:infinite}).

\begin{table}[b]
\caption{\label{tab:summary}Leading scaling exponents of the empirical mean $\langle L_{n} \rangle \sim n^{\theta}$ and variance $\langle L_{n}^{2} \rangle - \langle L_{n} \rangle^{2} \sim n^{2\gamma}$ of the length of the LIS of random walks with different distributions of step lengths. The values of $\alpha$ refer to the characteristic exponent of the symmetric $\alpha$-stable distributions, and the numbers between parentheses indicate the uncertainty in the last digit(s) of the data. The $\alpha$-stable distributions have infinite variance, while the uniform, Laplace and Gaussian distributions have all moments finite.}
\centering
\begin{tabular}{@{\hspace{.5em}}cccccccc} \hline\hline
${}$     & $\alpha=\frac{1}{2}$ & $\alpha=1$ & $\alpha=\frac{3}{2}$ & $\alpha=\frac{7}{4}$ & Uniform & Laplace & Gaussian \\ \hline
$\theta$ & $0.690(4)$ & $0.6851(3)$ & $0.6323(7)$ & 0.599(1) & $0.5680(15)$ & $0.568(2)$ & $0.567(2)$ \\
$\gamma$ & $0.704(1)$ & $0.6844(4)$ & $0.6347(6)$ & 0.601(1) & $0.568(2)$   & $0.568(2)$ & $0.568(2)$ \\ \hline\hline
\end{tabular}
\end{table}


\subsection{Scaling form}

\begin{figure}
\centering 
\includegraphics[viewport=0 10 480 460, scale=0.30, clip]{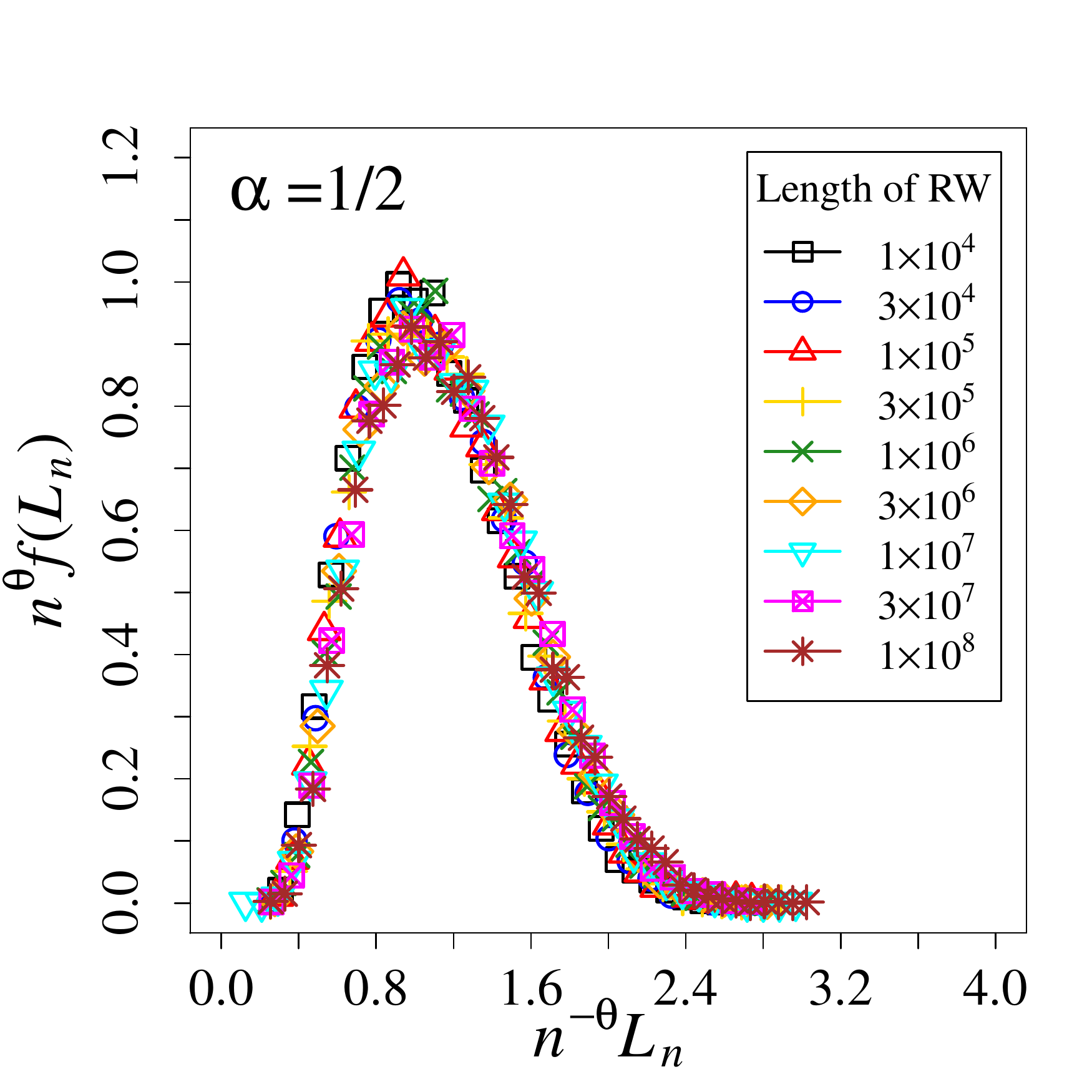} \hfill
\includegraphics[viewport=0 10 480 460, scale=0.30, clip]{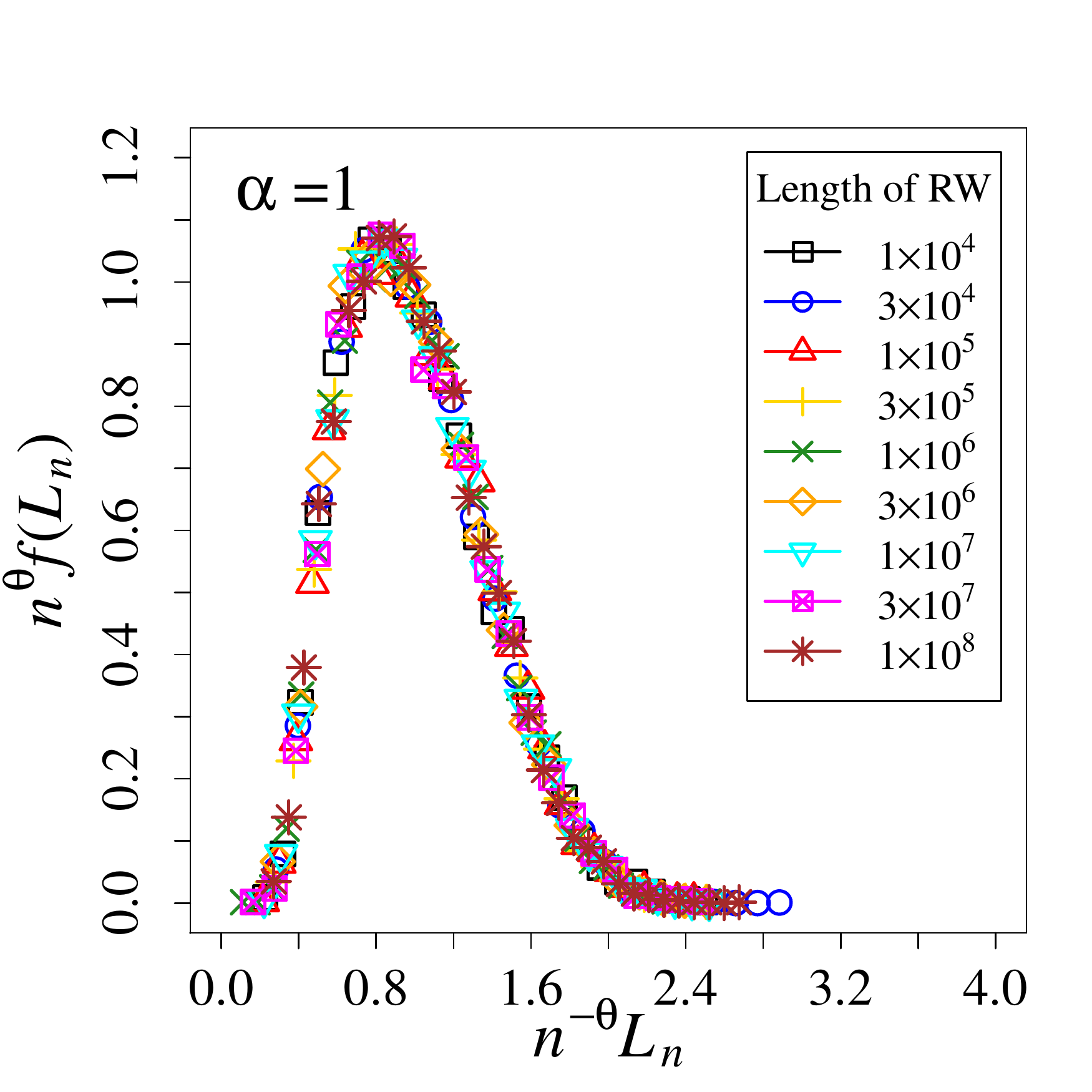} \hfill
\includegraphics[viewport=0 10 480 460, scale=0.30, clip]{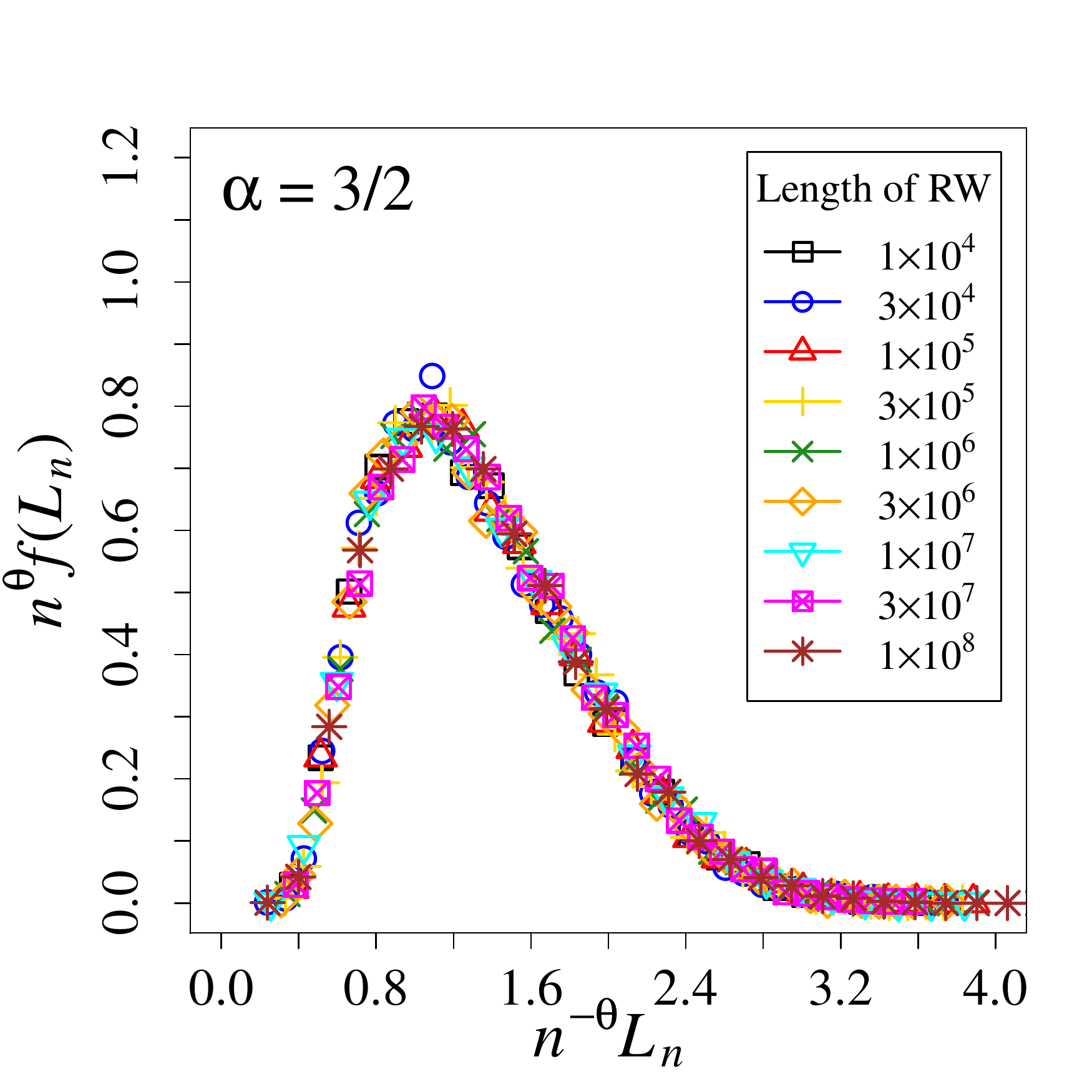} \\[1ex]
\includegraphics[viewport=0 10 480 460, scale=0.30, clip]{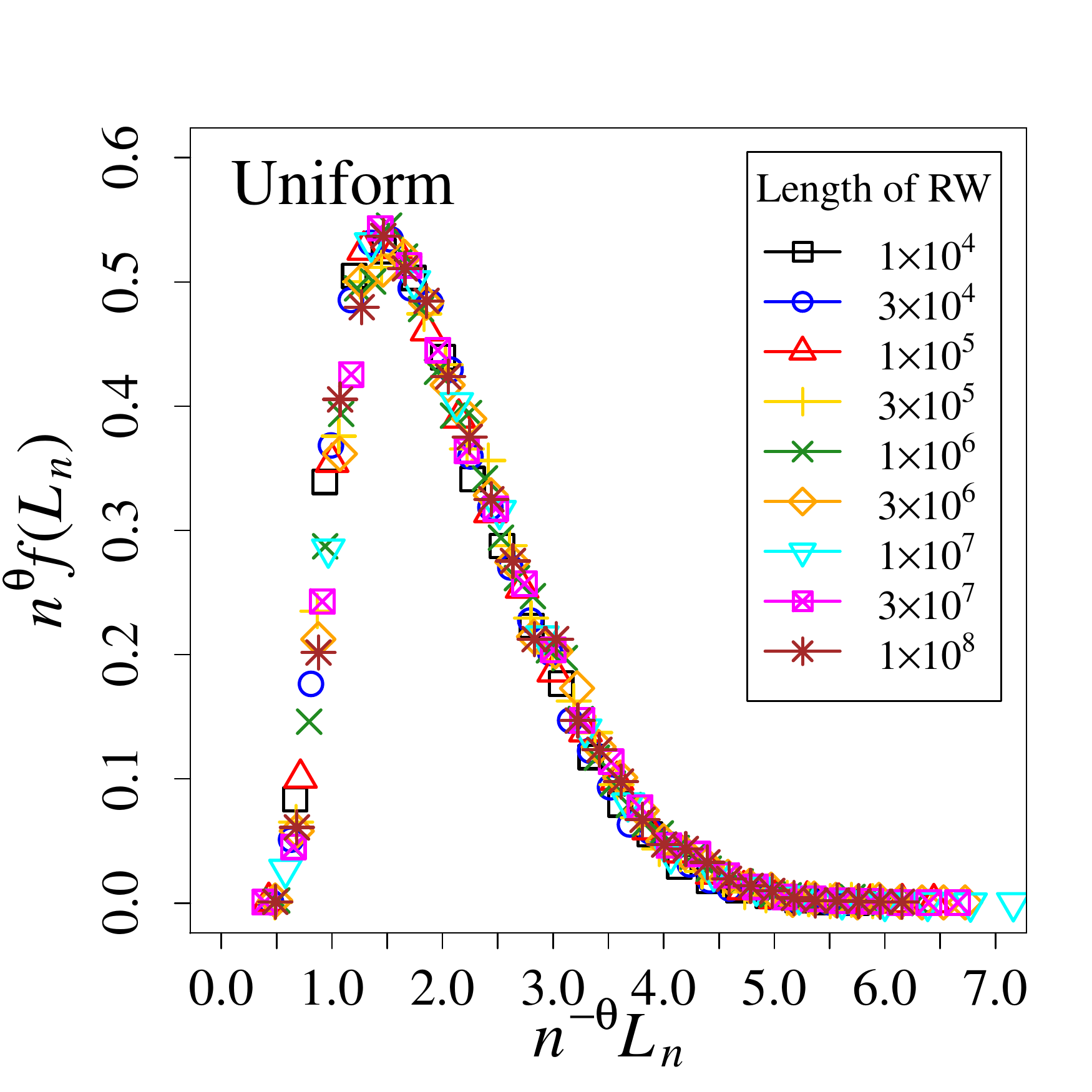} \hfill
\includegraphics[viewport=0 10 480 460, scale=0.30, clip]{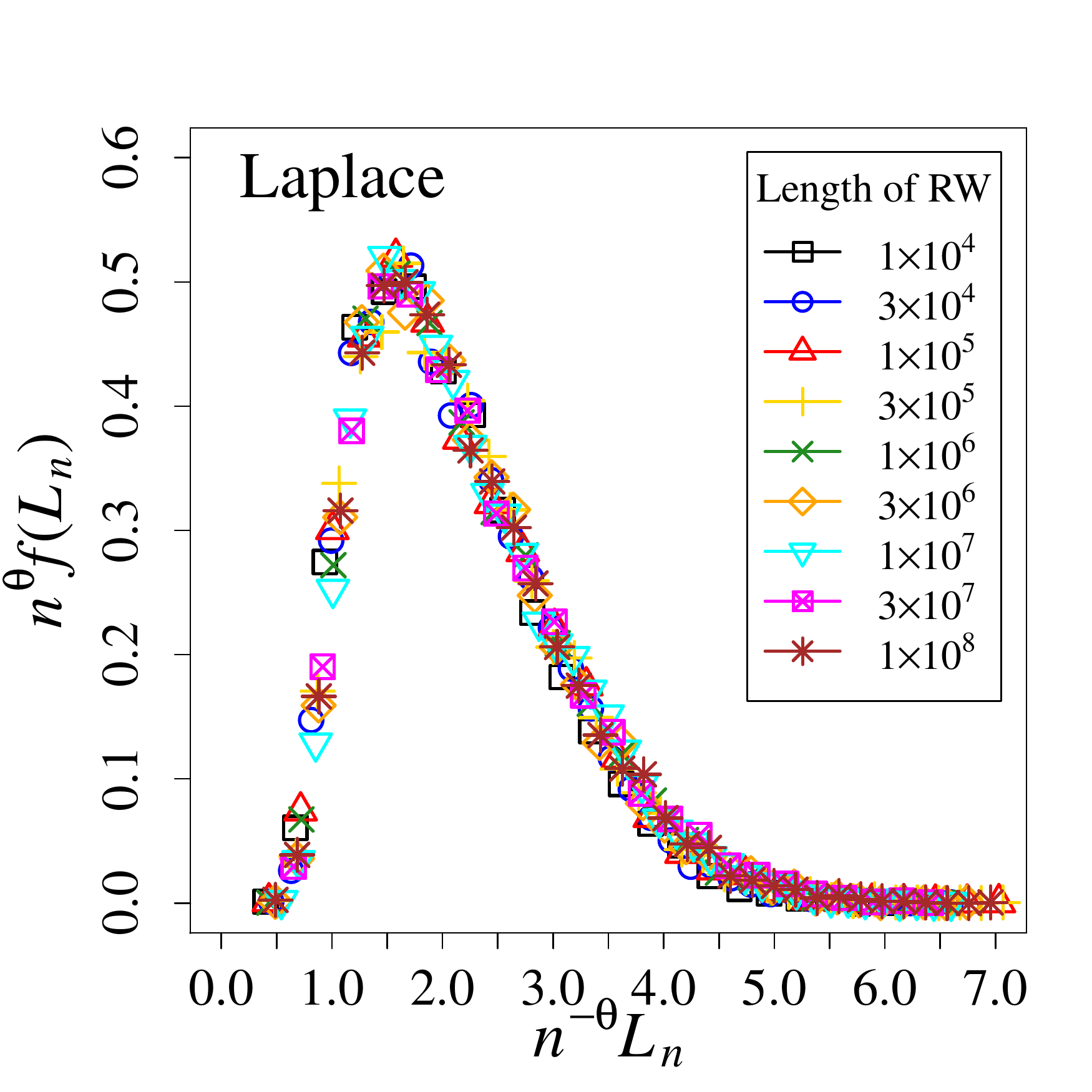} \hfill
\includegraphics[viewport=0 10 480 460, scale=0.30, clip]{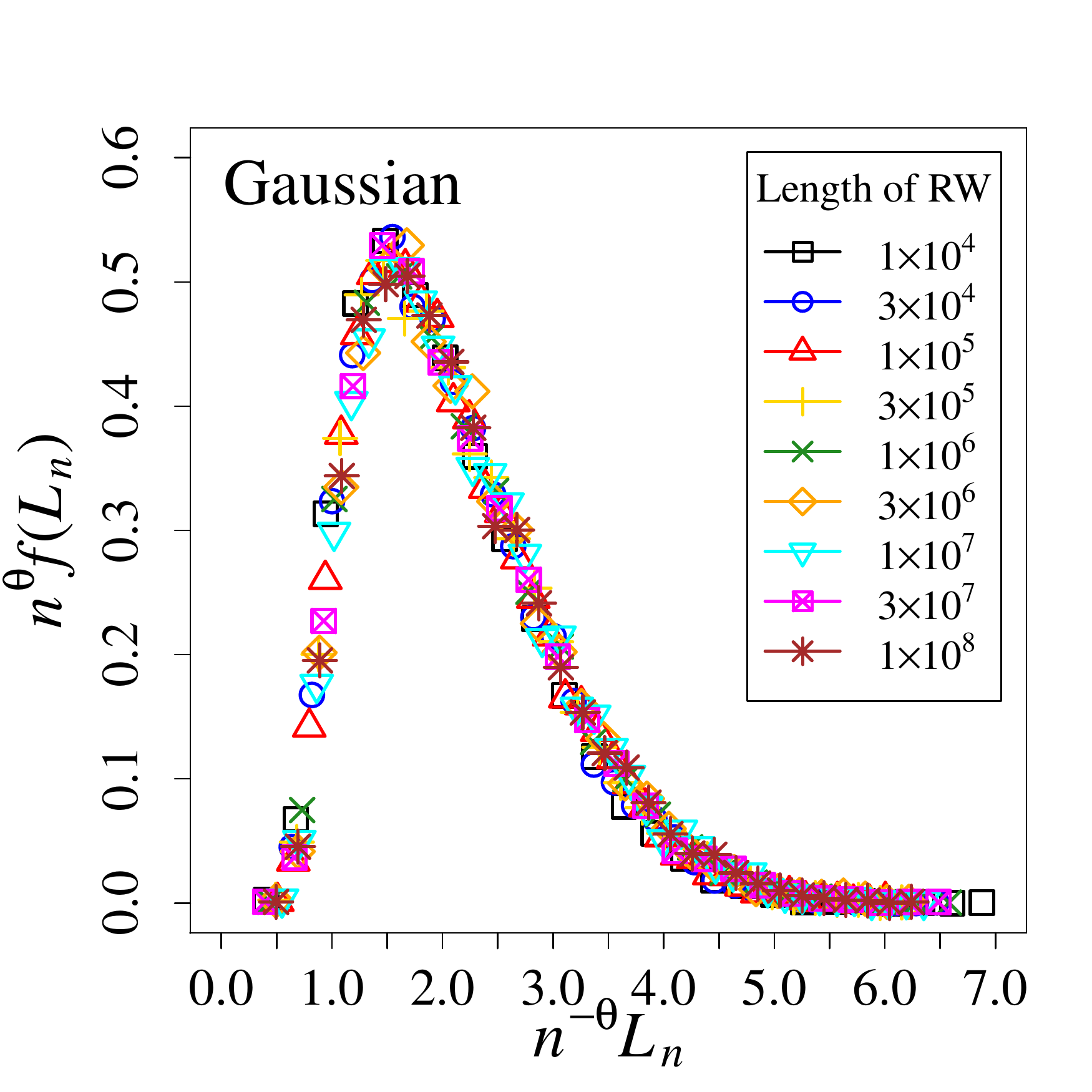}
\caption{\label{fig:collapse}Data collapse for the \pdf\ of the scale-adjusted $f(L_{n})$ according to (\ref{eq:scaling}) for some distributions of step lengths with infinite (upper panels) and finite (lower panels) variance. Note the different scales in the axes for the two different sets of figures. The scaling function for the $\alpha=\frac{7}{4}$ stable distribution (not shown) is shorter (maximum height $\sim 0.7$) and more spread out (until $\sim 5.5$) than the one for the $\alpha=\frac{3}{2}$ distribution, but not as much as the one for the Gaussian ($\alpha=2$) distribution, i.\,e., it `interpolates' between the two distributions.}
\end{figure}

The figures for $\theta$ and $\gamma$ in table~\ref{tab:summary} are virtually identical for all distributions examined, heavy-tailed or not. This evidence suggests that the \pdf\ of $L_{n}$ follows the simple scaling form
\begin{equation}
\label{eq:scaling}
f(L_{n}) = n^{-\theta}g(n^{-\theta}L_{n}),
\end{equation}
since then $\langle L_{n} \rangle \sim n^{\theta}$ and $\langle L_{n}^{2} \rangle - \langle L_{n} \rangle^{2} \sim n^{2\theta}$, as observed. To test this ansatz we plot $n^{\theta}f(L_{n})$ against $n^{-\theta}L_{n}$ to obtain $g(u)$. The data collapse observed in figure~\ref{fig:collapse} is clear for all distributions of step lengths, confirming (\ref{eq:scaling}) to a considerable degree.

The scaling functions seem to vary as the leading scaling exponent $\theta$ varies. Note, in particular, how the scaling functions for the $\alpha=\frac{1}{2}$ and $\alpha=1$ stable distributions look similar, as they beget LIS with similar exponents $\theta$, cf.~table~\ref{tab:summary}, despite their different tail behavior---i.\,e., the scaling function seems to depend more closely on $\theta$ than on the tail behavior of the underlying distribution of step lengths. Accordingly, the scaling functions for the distributions of step lengths of finite variance examined look universal, since they all share the same exponent $\theta$. We do not try to identify $g(u)$ in terms of known distributions; we intend to return to this issue in the future.

\subsection{\label{correction}Correction to scaling}

\begin{figure}
\centering
\includegraphics[viewport=0 10 480 460, scale=0.38, clip]{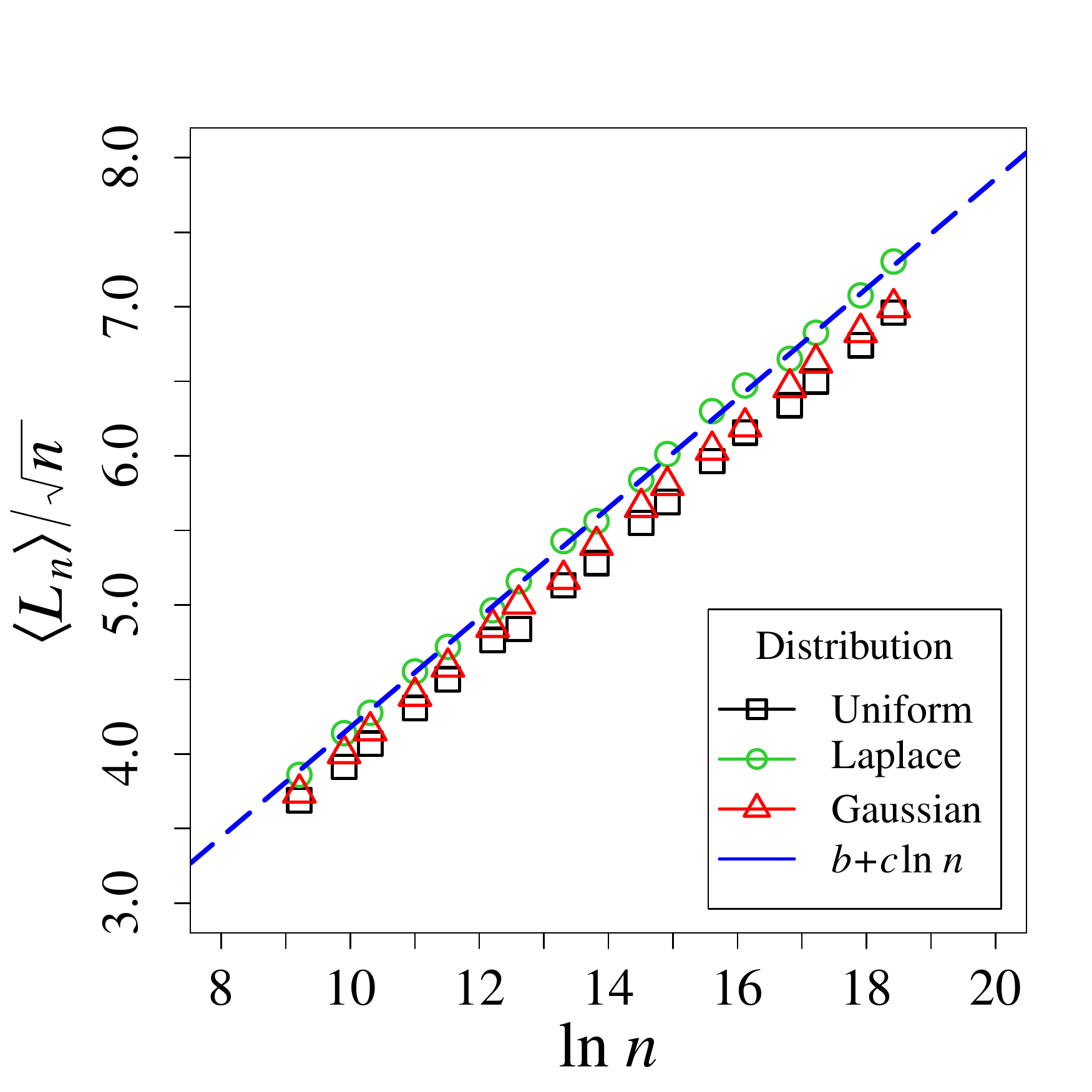} \hspace{1em}
\includegraphics[viewport=0 10 480 460, scale=0.38, clip]{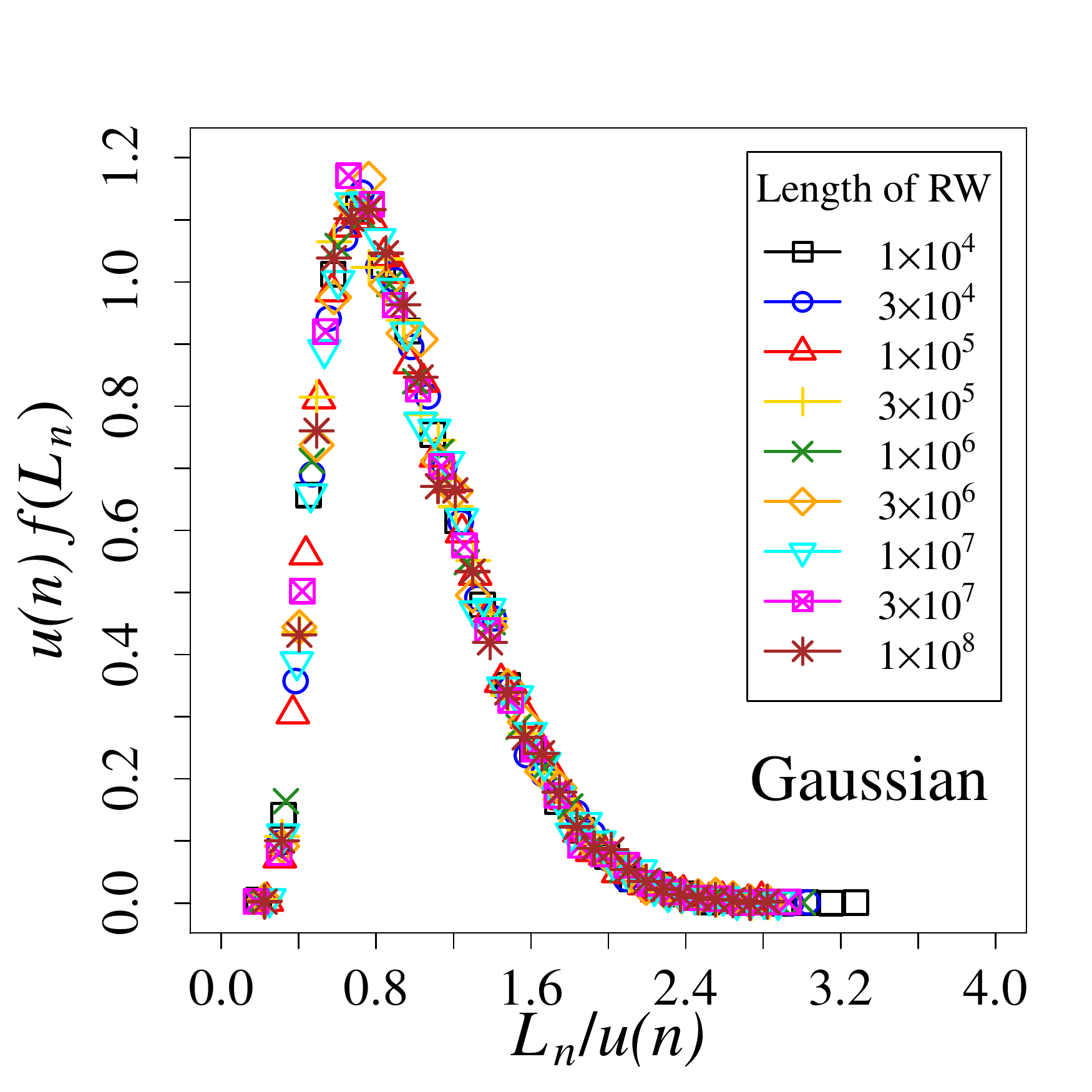}
\caption{\label{fig:correction}Left panel: Empirical mean $\langle L_{n} \rangle$ scaled by $\sqrt{n}$ against $\ln n$ for the uniform, Laplace and Gaussian distributions of step lengths together with the line $b+c\ln n$ (right-hand side of (\ref{eq:linear})) with $b=\frac{1}{2}$ and $c=\mathrm{e}^{-1}$ for comparison. Note the linear scale of the axes. Right panel: Data collapse for the \pdf\ of the scale-adjusted distribution of $L_{n}$ for the Gaussian random walk according to the conjectured form (\ref{eq:conjecture}), where $u(n)$ stands for the right-hand side of (\ref{eq:conjecture}).}
\end{figure}

The fits of $\langle L_{n} \rangle$ to $n^{\theta}$ are very good, and it would be difficult to assess corrections to this scaling directly from the data. However, for distributions of step lengths of finite variance, the bounds (\ref{eq:finite}) acknowledge that, besides the leading asymptotics $\langle L_{n} \rangle \sim \sqrt{n}$ (which we see from table~\ref{tab:summary} that definitely did not appear as such, at least not until $n=10^{8}$), there may be logarithmic corrections to the scaling. We thus test our data for the uniform, Laplace and Gaussian distributions to verify whether
\begin{equation}
\label{eq:correction}
{\langle L_{n} \rangle}/{\sqrt{n}} \sim (\ln n)^{a}
\end{equation}
for some positive constant $a$. A simple plot of ${\langle L_{n} \rangle}/\sqrt{n}$ against $\ln n$ for the three distributions appear in figure~\ref{fig:correction}. This figure reveals an intriguing linear relation
\begin{equation}
\label{eq:linear}
{\langle L_{n} \rangle}/{\sqrt{n}} \simeq b + c\ln n,
\end{equation}
with $b \simeq \frac{1}{2}$ and $c \simeq 0.36$ in all cases, the intriguing part being the constant $b$. This encouraged us to try to obtain a more precise estimate for the exponent $a$ by adjusting
\begin{equation}
\label{eq:found}
\ln\bigg(\frac{L_{n}-\frac{1}{2}\sqrt{n}}{\sqrt{n}}\bigg) \simeq \ln c + a\ln\ln n.
\end{equation}
We found that $\ln c \simeq -1$ and $a \simeq 1$ for all three distributions, cf.~table~\ref{tab:ablnc}.

\begin{table}[b]
\caption{\label{tab:ablnc}Constants appearing in (\ref{eq:linear}) and (\ref{eq:found}) for the distributions of step lengths examined.}
\centering
\begin{tabular}{@{\hspace{.5em}}cccc} \hline\hline
${}$    &   Uniform   &   Laplace    &   Gaussian   \\ \hline
$b$     &  $0.42(3)$  &  $0.46(3)$   &  $0.50(3)$   \\
$\ln c$ & $-1.10(1)$  & $-1.026(15)$ & $-1.043(14)$ \\
$a$     &  $1.018(5)$ &  $1.010(6)$  &  $1.002(5)$  \\ \hline\hline
\end{tabular}
\end{table}

Inspired by Hammersley, that commenting on \cite{ulam} stated that `papers are more en\-ter\-tain\-ing if they are still rich in conjectures, with results unproved or even wrong' \cite[p.~349]{hammersley}, we conjecture, based on the numerical evidence provided by (\ref{eq:linear}), (\ref{eq:found}), and table~\ref{tab:ablnc}, that the length $L_{n}$ of the LIS of random walks with step lengths of finite variance scales with $n$ like
\begin{equation}
\label{eq:conjecture}
L_{n} \sim \frac{1}{\mathrm{e}}\sqrt{n}\ln n + \frac{1}{2}\sqrt{n}
\end{equation}
plus lower order terms, although we concede that, as far as the constants go, it amounts to little more than numerology. Note that (\ref{eq:conjecture}) pushes the lower bound (\ref{eq:finite}) for the LIS of random walks of finite variance on $\mathbb{R}$ up by a factor of $\ln n$. Figure~\ref{fig:correction} (right panel) displays the \pdf\ of the scale-adjusted distribution of $L_{n}$ for the Gaussian random walk according to the conjectured form (\ref{eq:conjecture}). The scales of the axes in the figure now match the scales for the other distributions in figure~\ref{fig:collapse}. The very good data collapse and the coincident scales are repeated for the uniform and Laplace distributions of step lengths.


\section{\label{conclusion}Summary and outlook}

\subsection{\label{summary}Summary}

We found that the length of the LIS of random walks scales with the length of the walk as $L_{n} \sim n^{\theta}$ with an exponent that varies from $\theta=0.690(4)$ for an $\alpha=\frac{1}{2}$ stable distribution of step lengths down to $\theta=0.567(2)$ for the Gaussian ($\alpha=2$ stable) random walk, and that for the symmetric uniform and Laplace distributions the value of $\theta$ is the same as that for the Gaussian random walk. This indicates that while $\theta$ depends on the heaviness of the distribution of step lengths, for distributions of finite variance it is, barring logarithmic corrections, universal. For heavy-tailed distributions $\theta$ approaches the rigorous lower bound (\ref{eq:infinite}) from below. This is somewhat surprising, but not inconsistent with the rigorous bound, which was obtained for a `fat-tailed' distribution that is equivalent to an $\alpha \to 0$ stable distribution, and our results already for $\alpha=\frac{1}{2}$ fit within the bounds. Another noteworthy feature of the empirical exponents is that they are much closer to the rigorous lower bound ($\beta_{0} \simeq 0.690$) than to the upper bound ($\beta_{1} \simeq 0.815$), perhaps indicating that the techniques employed in \cite{pemantle} to obtain the lower bound capture better the nature of the LIS of random walks---unless $\theta$ varies more wildly as $\alpha \searrow 0$, which does not seem to be the case since from table~\ref{tab:summary} it seems that the second derivative $\theta{'}{'}\!(\alpha^{-1})<0$. It would be interesting to extend table~\ref{tab:summary} to include other heavy-tailed distributions of step lengths to better understand the dependence of $\theta$ on the heavy tails. For small values of $\alpha$, one can appeal to the Student-$t$ distribution of real $\nu > 0$ `degrees of freedom,' a symmetric distribution with tails decaying like $\abs{t}^{-\nu-1}$ for which random deviates can be efficiently and reliably generated \cite{bailey}. The ultra-heavy tail limit $\alpha \to 0$  can be emulated, for example, with the symmetric log-Cauchy random variable $R\mathrm{e}^{X}$ with $R$ a random sign and $X \sim \cauchy{\delta}{\gamma}$, which exhibits tails decaying like $\abs{x}^{-1}(\ln\abs{x})^{-2}$. The actual simulation of very heavy-tailed random walks is not without numerical subtleties, mainly because one needs to add numbers of very widely different orders of magnitude while keeping their full significance.

We also found that the empirical variance of the length of the LIS scales with the length of the random walk as $\langle L_{n}^{2} \rangle - \langle L_{n} \rangle^{2} \sim n^{2\gamma}$ with $\gamma$ virtually identical to $\theta$ for all distributions investigated. This suggested that the \pdf\ of $L_{n}$ follows the simple scaling form (\ref{eq:scaling}), which we confirmed by data-collapsing. It would be desirable to characterize the scaling functions $g(u)$ in (\ref{eq:scaling}) in more detail, in particular their tail behavior and whether they correspond to known distributions or can be expressed in terms of known functions.

The rigorous bounds for the scaling behavior of the LIS of random walks of finite variance acknowledge possible logarithmic corrections to (\ref{eq:finite}), and we found that our data are indeed compatible with a scaling behavior of the form $\sim\sqrt{n}\ln n$. This shows, nonrigorously, that the previous lower bound $\mathbb{E}(L_{n}') \geq c\sqrt{n}\ln n$ for the simple random walk on $\mathbb{Z}$, where $L_{n}'$ is the length of the weakly increasing subsequence (cf.~remarks following (\ref{eq:finite})) extends to random walks on $\mathbb{R}$. Incidentally, our analyses also suggested the form of the first correction to scaling, resulting in ours having conjectured---to make this paper more en\-ter\-tain\-ing---that for random walks with step lengths of finite variance $L_{n}$ scales like in (\ref{eq:conjecture}). Any proof, disproof, or cor\-rec\-tion to this conjectured asymptotics would be welcome.


\subsection{\label{outlook}Outlook}

We currently lack a working model, microscopic or hydrodynamic, for the length of the LIS of random walks that could allow for analytical approaches to its scaling behavior. We guess that some kind of renormalization of the subsequences of the random walk may bring some tractability to the problem. In a random walk of length $n$, after $k$ steps there remain $n-k$ steps that may contribute to the LIS of the walk. If $k$ is large (say, $k \gg n-k$), only paths that stay above the last element of the LIS have chance to contribute to the LIS. This is not exact, because if the LIS up to step $k$ is smaller than $n-k$ (and whether this is a rare event or not is relevant), than a `revolution' may occur and the LIS may become entirely contained in the second part of the walk. But this rationale suggests a sort of renormalization procedure and, in fact, some of the techniques employed in \cite{angel,pemantle} resemble such ideas.

The LIS problem for random walks brings to mind the problem of characterizing the records of random walks \cite{gumbel,sat-phys-a}. The LIS of a random walk, however, is a much more intricate quantity than its set of records, since it depends on the whole walk---the addition of a new term to the sequence does not change the set of records to date except for the occasional addition of a new record, but can change dramatically the associated LIS (the LIS itself, not its length, that can at most increase by $1$). This is clear also from an algorithmic perspective: while the set of records of a sequence of $n$ numbers is computable in $O(n)$ operations, the determination of one LIS of the same sequence is in general an $O(n\log_{2}n)$ operation \cite{fredman}. A similar phenomenon occurs with the time intervals between successive records (the ages of the records) of a symmetric random walk, which do not behave like independent \rv's and are sensitive to the last record \cite{sat-ziff,mounaix,god-sat-greg}. Whether there are connections between the statistics of records of random walks or of their ages---or of any other model or observable, for that matter \cite{spohn,nechaev}---and those of LIS of random walks is a question of considerable interest.


\section*{Acknowledgments}

The author is pleased to thank Satya N. Majumdar and Gr\'{e}gory Schehr for key observations on the scaling of the LIS of random walks and for kind hospitality during his visit (5--15 September 2016) to LPTMS/CNRS, Universit\'{e} Paris-Sud, Orsay, France. He also benefited from useful correspondence with Omer Angel (UBC), Robin Pemantle (UPenn), and Yuval Peres (Microsoft), to whom he is grateful. This work was partially supported by FAPESP, the S\~{a}o Paulo State Research Foundation, under grant 2015/21580-0.


\vspace{2ex}

\hspace*{\fill} $\star \quad \star \quad \star$ \hspace*{\fill}

\end{document}